\documentclass[aps,pra,twocolumn,groupedaddress,superscriptaddress,floatfix,showpacs]{revtex4-1}

\usepackage{amsmath, amsthm, amssymb}
\usepackage{graphicx}
\usepackage{psfrag}

\usepackage{xcolor}
\definecolor{mydarkgreen}{rgb}{0.0,0.5,0.0}
\usepackage[linktocpage=true,
  colorlinks=true, 
  pdfborder={0 0 0},
  linkcolor=blue,
  citecolor=red,
  filecolor=yellow,
  urlcolor=mydarkgreen,
  bookmarks,
  pdftitle={Foundations\ of\ FT-RDMFT},
  pdfauthor={Tim Baldsiefen},
  pdfkeywords={Reduced density matrix functional theory},
]{hyperref} 

\newcommand{\beal}{\begin{align}}
\newcommand{\eal}{\end{align}}
\newcommand{\tb}[1]{\textbf{#1}}
\newcommand{\tr}[1]{\textup{tr}\{#1\}}
\newcommand{\Eq}[1]{(\ref{#1})}
\newcommand{\Tr}[1]{{\textup{Tr}\left\{#1\right\}}}

\newcommand{\ket}[1]{{|#1\rangle}}
\newcommand{\bra}[1]{{\langle#1|}}

\newcommand{\nn}{\nonumber}
\newcommand{\veps}{\varepsilon}

\usepackage{feynmp}
\usepackage{simplewick} % Wick contractions

\newlength{\ltiny}
\newlength{\lsmall}
\newlength{\lmedium}
\newlength{\lbig}
\newlength{\lbigg}
\newlength{\lhuge}
\newcommand{\arrowshort}{7}
\newcommand{\arrowlong}{10}
\newcommand{\arrowang}{25}
\newcommand{\xtiny}{5}
\newcommand{\xsmall}{10}

\newcommand{\xbig}{20}
\newcommand{\xbigg}{30}
\newcommand{\xhuge}{40}

\def\sss{\scriptscriptstyle\rm}
\def\eq{_{\sss eq}}
\def\s{_{\sss S}}
\def\h{_{\sss H}}
\def\x{_{\sss X}}
\def\c{_{\sss C}}
\def\hx{_{\sss HX}}

\def\w{_{\sss W}}

\newlength{\miniwidth}
\newlength{\miniheight}
\newlength{\minicircwidth}
\begin{document}
\title{Reduced Density Matrix Functional Theory at Finite Temperature:\\ Theoretical Foundations}

\author{Tim Baldsiefen}
\affiliation{Max-Planck-Institut f\"ur Mikrostrukturphysik, Weinberg 2, D-06112 Halle, Germany}
\affiliation{Institut f\"ur Theoretische Physik, Freie Universit\"at Berlin, Arnimallee 14, D-14195 Berlin, Germany}
\author{Attila Cangi}
\affiliation{Max-Planck-Institut f\"ur Mikrostrukturphysik, Weinberg 2, D-06112 Halle, Germany}
\author{E.K.U. Gross}
\affiliation{Max-Planck-Institut f\"ur Mikrostrukturphysik, Weinberg 2, D-06112 Halle, Germany}

\date{\today}

\begin{abstract}
  We present an ab-initio approach for grand canonical ensembles in
  thermal equilibrium with local or nonlocal external potentials 
  based on the one-reduced density matrix. 
  We show that equilibrium properties of a grand canonical ensemble 
  are determined uniquely by the eq-1RDM and establish a variational principle 
  for the grand potential with respect to its one-reduced
  density matrix. We further prove the existence of a Kohn-Sham system 
  capable of reproducing the one-reduced density matrix of an interacting system 
  at finite temperature. Utilizing this Kohn-Sham system as an unperturbed system, 
  we deduce a many-body approach to iteratively construct approximations to the 
  correlation contribution of the grand potential.
\end{abstract}


\maketitle

\section{Introduction\label{eq.intro}}
Based on the celebrated theorems of Hohenberg and Kohn\cite{HK64}, 
Kohn-Sham density functional theory (KS-DFT)\cite{KS65} 
is currently the method of choice for calculating groundstate (gs) properties 
of quantum systems.

There are, however, cases in which KS-DFT performs rather poorly. 
A prominent example is its failure in predicting the fundamental gap, 
in particular, of so-called Mott insulators\cite{GMR06,GSSG11}.
KS-DFT with standard exchange-correlation approximations fails 
for this kind of strongly correlated system 
and typically yields a metallic ground state, 
while the true experimental low-temperature phase is 
that of an antiferromagnetic insulator.
At finite temperature the description of strongly correlated systems is even
more challenging. 
Genuine Mott insulators exhibit a characteristic feature: 
when heated up from their antiferromagnetic insulating gs, they stay
insulating above the N\'{e}el temperature, i.e., in the absence of long-range
magnetic order. Contrarily, weakly correlated insulators 
-- so-called Slater insulators -- become metallic at the N\'{e}el temperature.

A possible approach to tackle this challenge is to search for more accurate
functionals in the framework of KS-DFT. Alternatively, one may look for other
theoretical frameworks in which the treatment of strong correlation might be
simpler. One candidate for such a framework is reduced density matrix functional
theory (RDMFT). Through its more direct treatment of many-particle correlations
it has promising potential for calculations of finite\cite{B01,BB02,HLAG07,ML08,PMLU10,HLG09}
as well as infinite systems\cite{LSHD10,LSDF09,SDLG08}. In particular, it
was possible to predict insulating ground states for transition metal oxides
without breaking the spin symmetry\cite{SDLG08}.

Motivated by the success of RDMFT at zero temperature, 
the purpose of the present work is to lay the theoretical foundations 
for the finite-temperature version of RDMFT (FT-RDMFT). 
As a general ab-initio theory its applicability is not restricted 
to Mott insulators. 
There is a variety of physical phenomena, 
in particular in the warm dense matter regime\cite{GDRT14}  
which requires an accurate description of quantum effects at finite temperature\cite{KD09}.
These phenomena include temperature-driven magnetic\cite{L79,RJ97} 
or superconducting\cite{PFLF06,CPSF08} phase transitions in solids, 
femtochemistry at surfaces of solids\cite{GROT09}, 
properties of shock compressed noble gases\cite{M06,RMCH10},
the properties of plasmas\cite{DP82,PD00,DM08}, 
thermal conductivities of inertial confinement fusion capsules\cite{A04},
and planetary interiors and their formation processes\cite{NBHR12,NHFR13,LHR09,LHR11,RMNF11}.

This paper is divided as follows:
In Sec.~\ref{sec.FT-RDMFT} we derive and present the formalism of FT-RDMFT.
First, in Sec.~\ref{sec.FT-RDMFT.background} we introduce our notation. Note 
that we work in atomic units throughout where $e^2=\hbar=m_e=1$ so that
lenghts are expressed in Bohr radii, and energies in hartree.
Then, in Sec.~\ref{sec.FT-RDMFT.hk} we lay the foundations of FT-RDMFT 
by showing that the grand potential of systems with generally nonlocal external potentials 
can be written as a functional of the one-reduced density matrix (1RDM).
Next, in Sec.~\ref{sec.FT-RDMFT.ks} we show the existence of a KS system 
in FT-RDMFT and demonstrate how the KS Hamiltonian is explicitly constructed.
Subsequently, in Sec.~\ref{sec.FT-RDMFT.ac} we derive the adiabatic connection formula
which forms the basis for the construction of approximations to the correlation functional
in FT-RDMFT.
Finally, in Sec.~\ref{sec.FT-RDMFT.functionals} 
the existence of a KS system and the adiabatic connection formula 
enable us to derive a methodology for iteratively constructing correlation functionals 
based on finite-temperature many-body perturbation theory (FT-MBPT). 
Furthermore, in Appendix~\ref{sec.appendix.eq-occ} we give a detailed analysis of occupation
numbers in interacting systems, 
in Appendix~\ref{sec.appendix.counter} we investigate the one-to-one mapping 
between the external potential and the wavefunction at zero temperature,
in Appendix~\ref{sec.appendix.FT-HF} we show that our iterative procedure for constructing
functionals from FT-MBPT yields the finite-temperature Hartree-Fock functional 
as the first-order contribution,
and in Appendix~\ref{sec.appendix.canonical} we present the formulation of FT-RDMFT 
for a canonical ensemble.

\section{Finite-temperature reduced density matrix functional theory}
\label{sec.FT-RDMFT}

\subsection{Background}
\label{sec.FT-RDMFT.background}

The main thermodynamic variable in a grand canonical ensemble is the grand
potential
\begin{align}
\label{eq.def.gp}
\Omega[\hat D]&=\tr{\hat D \hat \Omega} 
\end{align}
given as a statistical average over the grand canonical operator
\begin{align}
\hat \Omega = \hat H + \mu \hat N + \hat S/\beta\,,
\end{align}
where $\hat H$, $\hat N$, and $\hat S$ are the Hamiltonian, particle number, 
and entropy operators. In electronic structure theory the Hamiltonian is typically 
given by $\hat H = \hat T + \hat W + \hat V$, where $\hat T$ denotes the kinetic energy operator,
$\hat W$ the interelectronic repulsion in a Coulombic system, 
and $\hat V$ represents a scalar external potential.
The coupling to particle and heat baths is achieved via the Lagrangian multipliers 
$\mu$ denoting the chemical potential and $1/\beta$ denoting the temperature.

Statistical averages as in Eq.~(\ref{eq.def.gp}) are computed via 
the statistical density operator (SDO) $\hat D$ which is defined as 
a weighted sum of projection operators on the underlying Hilbert space.
The appropriate Hilbert space for grand canonical ensembles, where a change of particle
number is allowed, is a direct sum of symmetrized
tensor products of the one-particle Hilbert space -- called the Fock space.
Assuming that the system does not allow for mixing of states with different particle numbers, 
the set of all possible SDOs can be expressed just by
projection operators on states with defined particle number $N$:
\begin{multline}
\hat D=\sum_{\alpha,N}w_{\alpha N}\ket{\Psi_{\alpha N}}\bra{\Psi_{\alpha N}},\\w_{\alpha N}\geq0,\sum_{\alpha, N}w_{\alpha N}=1\label{eq.sdo.gc}\,, 
\end{multline}
where $\ket{\Psi_{\alpha N}}$ and $w_{\alpha N}$ are orthonormal $N$-particle states
and their corresponding weights.

The thermal equilibrium (eq) of a grand canonical ensemble is then defined as
that SDO for which the grand potential $\Omega[\hat D]$ is minimal. 
This definition leads to the finite-temperature Rayleigh-Ritz variational 
principle\cite{M63} which states that
\begin{align}
  \Omega[\hat D]&>\Omega[\hat D\eq],\quad \hat D\neq\hat D\eq\label{eq.varpr.ft},
\end{align}
with
\begin{align}
    \hat D\eq&=e^{-\beta(\hat H-\mu\hat N)}/\tr{e^{-\beta(\hat H-\mu\hat N)}}.\label{eq.deq}
\end{align}

The 1RDM is defined by the SDO and the help of the common fermionic field operators as
\begin{align}
  \gamma(x,x')&=\tr{\hat D\hat\psi^+(x')\hat\psi(x)}\label{eq.def.1rdm}\,,
\end{align}
where the variable $x$ denotes a combination of spin index $\sigma$ and spatial coordinate $\tb
r$ ($x=(\tb r,\sigma)$). An integration over $x$ is therefore to be interpreted as an
integration over $\tb r$ and a summation over $\sigma$. 
Since the 1RDM is hermitean by construction, it is commonly written in spectral representation as
\begin{align}
  \gamma(x,x')&=\sum_in_i\phi_i^*(x')\phi_i(x)\label{eq.def.1rdm.2} 
\end{align}
with real-valued eigenvalues $\{n_i\}$ and eigenfunctions $\{\phi_i(x)\}$,
which are called occupation numbers and natural orbitals\cite{L55}.
The neccessary and sufficient conditions for N-representability\cite{C63} of $\gamma(x,x')$ 
are that $\{\phi_i\}$ is a complete set and
\begin{align}\label{eq.enr.2}
  0 \leq n_i \leq 1\,, \quad \sum_i n_i = N \ .
\end{align}

It is sometimes desirable to treat spin and spatial variables seperately. To this end we
introduce a two-component (Pauli) spinor notation.
\begin{align}
  \Phi_{i}(\tb r)=\left(    
    \begin{array}{c}
      \phi_{i1}(\tb r)\\
      \phi_{i2}(\tb r)
    \end{array}\right),
\end{align}
where $\phi_{i\sigma}(\tb r)=\phi_i(x)=\phi_i(\sigma,\tb r)\ (\sigma=1,2)$ are the
orbitals of Eq.~\Eq{eq.def.1rdm.2}. The 1RDM can then be written as a matrix in spin space
as
\begin{align}
  \gamma(\tb r,\tb r')&=\sum_in_i\Phi_i^\dagger(\tb r')\otimes\Phi_i(\tb r)\\
  &=\sum_in_i
  \left(
    \begin{array}{cc}
      \phi_{i1}^*(\tb r')\phi_{i1}(\tb r)&\phi_{i2}^*(\tb r')\phi_{i1}(\tb r)\\
      \phi_{i1}^*(\tb r')\phi_{i2}(\tb r)&\phi_{i2}^*(\tb r')\phi_{i2}(\tb r)
    \end{array}
  \right)\ .\label{eq.rdm.22}
\end{align}
In the special case of collinear spin configuration different spin channels can be treated seperately. 
For these systems, the natural orbitals are so-called spin orbitals, i.e.,
spinors containing only one spin component, where
\begin{align}
  \Phi_{i1}(\tb r)=\left(
  \begin{array}{c}
    \phi_{i1}(\tb r)\\
    0
  \end{array}\right)\,,\
  \Phi_{i2}(\tb r)=\left(
  \begin{array}{c}
    0\\
    \phi_{i2}(\tb r)
  \end{array}\right)\ .\label{eq.1rdm.no.special}
\end{align}
This leads to a 1RDM which has only one nonvanishing entry in every 2x2 matrix 
of Eq.~\Eq{eq.rdm.22}, either the 11 or the 22 element. 
Hence the complete 1RDM is diagonal
with respect to the spin coordinate
\begin{align}
  \gamma_{\sigma\sigma'}(\tb r,\tb r')&=\delta_{\sigma\sigma'}\sum_i n_{i\sigma}\phi^*_{i\sigma}(\tb r')\phi_{i\sigma}(\tb r),
\end{align}
where $n_{i\sigma}$ are the occupation numbers of the special spinors $\Phi_{i\sigma}(\tb
r)$ in Eq.~\Eq{eq.1rdm.no.special}. Spin spiral states are another special case 
where this seperation also applies\cite{BECG15}.

\subsection{Hohenberg-Kohn theorems for finite-temperature reduced density matrix functional theory}
\label{sec.FT-RDMFT.hk}

Mermin's extension of the HK theorems to finite temperature\cite{M65} 
immediately implies the one-to-one mappings
\begin{subequations}
\begin{align}
&\gamma(\tb r,\tb r') \longrightarrow 
\rho(\tb r) \stackrel{HK}{\longleftrightarrow} v(\tb r){\rm ~~and~} \label{eq.proof_1}\\
&v(\tb r) \longrightarrow 
\Psi(\hat D) \longrightarrow \gamma(\tb r,\tb r')\,, \label{eq.proof_2}
\end{align}
\end{subequations}
i.e., between the 1RDM, the density, the eq-SDO, 
and the \emph{local} external potential\footnote{Every 1RDM yields 
the density as its diagonal and the HK or Mermin theorems
prove the one-to-one correspondence between the density 
and the potential (see Eq.~\Eq{eq.proof_1}). 
This potential, on the other hand, determines the Hamiltonian and
therefore the groundstate $\Psi$ as well as the eq-SDO $\hat D$ which in turn lead to the
1RDM via Eq.~\Eq{eq.def.1rdm}. Therefore, Eqs.~\Eq{eq.proof_1} and \Eq{eq.proof_2}
are closed, and the HK and Mermin theorems imply the one-to-one correspondences
in Eqs.~(\ref{eq.proof_1}) and (\ref{eq.proof_2}).}.

However, in FT-RDMFT we need to go further than this and consider 
\emph{nonlocal} external potentials, in which case the ground state 
is not uniquely determined by the density anymore, but by the 1RDM\cite{G75}. 
Going beyond local external potentials is necessary, because 
the KS potentials in FT-RDMFT are \emph{nonlocal} in general, 
as we show in Sec.~\ref{sec.FT-RDMFT.ks}.

Therefore, we lay the foundations of FT-RDMFT and establish the uniqueness
of the KS scheme in FT-RDMFT by proving the one-to-one correspondences in 
Eqs.~(\ref{eq.proof_1}) and (\ref{eq.proof_2}) for nonlocal external potentials
at finite temperature. We also show that in this case the 1RDM is still sufficient to
describe equilibrium properties.
We divide this up into three steps, namely showing 
(i) that the map between $\hat D\eq$ and $\gamma\eq(x,x')$ has to be invertible, 
implying the existence of a grand potential functional $\Omega[\gamma]$, 
(ii) the existence of a universal functional $F[\gamma]$ and 
(iii) that the minimization of $\Omega[\gamma]$ leads to the eq-1RDM. 
Note that we consider only eq-1RDMs for the proof in step (i). However, we can 
relax this restriction in step (ii).

\subsubsection*{\textbf{\textup{(i) Proof of 
$\hat D\eq\stackrel{1-1}{\longleftrightarrow} \gamma\eq(x,x')$, 
i.e., one-to-one mapping between eq-SDO and eq-1RDM}}}

We divide proving the existence of a one-to-one mapping between
the eq-SDO and the eq-1RDM into two parts:
\begin{align}\label{proof.1-1.eqSDO-gamma}
& \hat D\eq 
\stackrel{1-1}{\longleftrightarrow} 
\mu(x,x')
\stackrel{1-1}{\longleftrightarrow}
\gamma\eq(x,x')\,,
\end{align}
where we prove 
(i.i) the one-to-one mapping between $\hat D\eq$ and the nonlocal 
chemical potential $\mu(x,x')=v(x,x')-\mu$ and 
(i.ii) the one-to-one mapping between $\mu(x,x')$ and $\gamma\eq(x,x')$.

\paragraph*{\textbf{\textup{(i.i) Proof of 
$\hat D\eq\stackrel{1-1}{\longleftrightarrow} \mu(x,x')$,
i.e., one-to-one mapping between eq-SDO and nonlocal chemical potential.~ ~}}}
We show this with a \emph{proof by contradiction}. 
Let $\hat H$ and $\hat H'$ be two different Hamiltonians 
and assume they lead to the same SDO $\hat D$, where $\hat H'$ shall
differ from $\hat H$ only by a one-particle potential contribution $\Delta \hat V$. 
With Eq.~\Eq{eq.deq} this reads
\begin{align}
e^{-\beta(\hat H-\mu\hat N)}/Z 
=e^{-\beta(\hat H+\Delta \hat V-\mu\hat N)}/Z'\label{eq.proof_rel}\,,
\end{align}
where $Z$ and $Z'$ are the partition functions, e.g., $Z=\tr{e^{-\beta(\hat H-\mu\hat
N)}}$. Solving Eq.~\Eq{eq.proof_rel} for $\Delta \hat V$ yields
\begin{align}
\Delta \hat V 
&=\int dxdx'\Delta v(x',x)\hat\psi^+(x')\hat\psi(x)=-\frac1\beta\ln\frac{Z}{Z'}\label{eq.upot}.
\end{align}
We now need to show that a one-particle potential $\Delta v(x,x')\neq 0$ 
fulfilling this equality 
cannot exist, thereby contradicting our initial assumption. 
To proceed, we assume three different Slater determinants 
$\ket{X_1}=\ket{1,0,0,...},\ket{X_2}=\ket{0,1,0,...}$, and
$\ket{X_3}=\ket{1,1,0,...}$ in the basis $\{\chi_i\}$. 
The potential in this basis is denoted by 
$\Delta v_{ij}=\int dxdx' \Delta v(x',x)\chi_i^*(x')\chi_j(x)$. 
Calculating the expectation value of both sides of Eq.~\Eq{eq.upot} 
with respect to these three Slater determinants yields the following system of equations
\begin{align}\label{eq.proof.uu}
-\frac1\beta\ln\frac{Z}{Z'} 
&= \Delta v_{11} = \Delta v_{22} = \Delta v_{11} + \Delta v_{22} . 
\end{align}
This can only be fulfilled by $\Delta v_{11} = \Delta v_{22}=0$ and $Z=Z'$. 
A repetition of this argument for all possible bases then shows that 
only $\Delta \hat V=0$ fulfills Eq.~\Eq{eq.upot} which in
turn proves the one-to-one correspondence between $\hat D\eq$ and $v(x,x')$,
and hence the one-to-one correspondence 
between $\hat D\eq$ and $\mu(x,x')$.

This proof is valid for any finite temperature. It is based on the bijectivity of the
exponential function which allows us to invert Eq.~\Eq{eq.proof_rel}, leading to 
Eq.~\Eq{eq.proof.uu}. At zero temperature, however, this bijectivity breaks down.
Further elaborations on zero-temperature mappings 
between external potentials and wavefunctions are given in
Appendix~\ref{sec.appendix.counter}.

\paragraph*{\textbf{\textup{(i.ii) Proof of 
$\mu(x,x') \stackrel{1-1}{\longleftrightarrow} \gamma\eq(x,x')$,
i.e., one-to-one mapping between nonlocal chemical potential and eq-1RDM.~ ~}}}

In order to prove the one-to-one correspondence between 
$\mu(x,x')$ and $\gamma\eq(x,x')$ we use a \emph{proof by contradiction} again. 
Consider two Hamiltonians $\hat H$ and $\hat H'$ differing only in their external 
and chemical potentials. The corresponding grand potentials are given by
\begin{align}
\Omega[\hat D\eq]&=\tr{\hat D\eq(\hat H-\mu\hat N+1/\beta\ln{\hat D\eq})}\textrm{~ ~and}\\
\Omega'[\hat D'\eq]&=\tr{\hat D'\eq(\hat H'-\mu'\hat N+1/\beta\ln{\hat D'\eq})},
\end{align}
where $\hat D\eq$ and $\hat D'\eq$ are defined by Eq.~\Eq{eq.deq}. 
Using $\hat D\eq\neq\hat D\eq'$ as we have proven in (i.i), 
the variational principle in Eq.~\Eq{eq.varpr.ft} then leads to
\begin{align}
\Omega[\hat D\eq]&<\Omega[\hat D'\eq]\\
&=\tr{\hat D'\eq(\hat H-\mu\hat N+1/\beta\ln{\hat D'\eq})}\\
&=\Omega'[\hat D'\eq]+\tr{\hat D'\eq((\hat H-\mu\hat N)-(\hat H'-\mu'\hat N))}\ .
\end{align}
By exchanging primed and unprimed quantities we obtain
\begin{align}
\Omega[\hat D\eq]&<\Omega'[\hat D'\eq]+\int dxdx'(\mu(x',x)-\mu'(x',x))\gamma'(x,x')\label{eq.rel.1}\\
\Omega'[\hat D'\eq]&<\Omega[\hat D\eq]+\int dxdx'(\mu'(x',x)-\mu(x',x))\gamma(x,x')\label{eq.rel.2}\ .
\end{align}
Adding these two equations leads to
\begin{align}
  \int dxdx'(\mu'(x',x)-\mu(x',x))(\gamma(x,x')-\gamma'(x,x'))&>0\label{eq.rel.3}.
\end{align}
The existence of two different sets of external and chemical potentials yielding the same
eq-1RDM lets the integral in Eq.~\Eq{eq.rel.3} vanish which leads to a contradiction. Hence
the initial assumption is falsified. 
This concludes our proof of Eq.~(\ref{proof.1-1.eqSDO-gamma}).

Having established the existence of a one-to-one mapping between $\hat D\eq$ and
$\gamma\eq$ we can now proceed and properly define the grand potential 
as a functional of the 1RDM as
\begin{align}\label{eq.gp.0}
\Omega[\gamma\eq] 
&=\tr{\hat D[\gamma\eq](\hat H-\mu\hat N+1/\beta\ln(\hat D[\gamma\eq]))}\ .
\end{align}

\subsubsection*{\textbf{\textup{(ii) Existence of a universal functional 
$F[\gamma\eq]$}}}

In analogy to DFT, we define a universal functional by separating 
the external and the chemical potential contributions from Eq.~\Eq{eq.gp.0}:
\begin{align}\label{F.gamma_eq}
F[\gamma\eq]&=\tr{\hat D[\gamma\eq](\hat T+\hat W+1/\beta\ln\hat D[\gamma\eq])}
\end{align}
such that 
\begin{align}
\Omega[\gamma\eq]&=\int dxdx' \mu(x,x')\gamma\eq(x,x')+\mathcal{F}[\gamma\eq]\ .
\end{align}
Notice a subtlety involved with defining the universal functional 
in Eq.~(\ref{F.gamma_eq}). In our proof we considered
a restricted set of 1RDMs, namely those coming from eq-SDOs given by Eq.~\Eq{eq.deq}. 
However, the conditions to ensure that an arbitrary 1RDM comes from such a SDO are unknown. 
Nevertheless, following ideas of Valone\cite{V80} and Lieb\cite{L83} 
we can resolve this subtle point and extend the domain of $\Omega[\gamma]$ 
to the whole set of ensemble-N-representable 1RDMs.
Accordingly we can now define the universal functional as
\begin{align}
\mathfrak{F}[\gamma] 
&=\inf_{\hat D\rightarrow\gamma}\tr{\hat D(\hat T+\hat W+1/\beta\ln\hat D)}
\end{align}
such that 
\begin{align}
\Omega[\gamma] 
&=\int dxdx' \mu(x,x')\gamma(x,x')+\mathfrak{F}[\gamma] .
\end{align}

\subsubsection*{\textbf{\textup{(iii) Minimization of $\Omega[\gamma]$}}}

The variational principle in Eq.~\Eq{eq.varpr.ft} now allows us to 
determine the equilibrium grand potential by 
\begin{align}
\Omega\eq&=\min_{\gamma\in\Gamma^N}\Omega[\gamma]\,,
\end{align}
a minimization over $\Gamma^N$ 
which is the set of all ensemble-N-representable 1RDMs. 
Additionally, we postulate 
\begin{align}\label{eq.euler}
\left.\frac{\delta \mathfrak{F}[\gamma]}{\delta\gamma(x,x')}\right|_{\gamma\eq}+v(x,x')
&=\mu\,,
\end{align}
the Euler-Lagrange equation for the eq-1RDM 
in FT-RDMFT\footnote{There are two subleties with Eq.~(\ref{eq.euler}): 
(i) the existence of the functional derivative 
$\delta \mathfrak{F}[\gamma]/\delta\gamma(x,x')|_{\gamma\eq}$ 
has not been proven so far and is subject to continued studies 
(also in DFT this question is still under discussion\cite{L07}). 
In an approximate treatment, however, one will most
likely define a functional $F^{approx}[\gamma]$ for which the functional derivative
exists, justifying the use of the Euler-Lagrange equation; 
(ii) if the eq-1RDM is on the boundary of the domain $\Gamma^N$, 
the Euler-Lagrange equation does not take the form of Eq.~\Eq{eq.euler} 
but has to incorporate constraints on the eigenvalues of the
1RDM by further Kuhn-Tucker multipliers\cite{KT51}. Fortunately, as we 
show in Appendix~\ref{sec.appendix.eq-occ}, 
the minimizing 1RDM will never be on the boundary of $\Gamma^N$ 
for grand canonical ensembles, rendering these additional multipliers unnecessary.}.

\subsection{Kohn-Sham system for finite-temperature reduced density matrix functional theory}
\label{sec.FT-RDMFT.ks}

We have established the theoretical framework of FT-RDFMT by proving Hohenberg-Kohn-like theorems. 
The central problem for turning this theory into a practical scheme is finding 
approximations as a functional of the 1RDM. In analogy to DFT, one possible route 
for constructing such approximations requires us to introduce the KS scheme.
Then we can exploit the existence of a KS system to derive a methodology 
for the iterative construction of functionals using methods from FT-MBPT.

Our starting point is an auxiliary system of noninteracting fermions described by the Hamiltonian
\begin{align}\label{KS.Ham}
\hat H^{(1)}\s = \hat T + \hat V\s\,, \quad 
\hat H^{(1)}\s&=\sum_i\varepsilon_i\ket{\phi_i}\bra{\phi_i}
\end{align}
with eigenvalues $\{\varepsilon_i\}$ and eigenfunctions $\{\phi_i(x)\}$, 
$\hat V\s$ denoting the operator of the KS potential.
Then we assume the existence of a nonlocal potential -- the KS potential $v\s(x,x')$ --  
which yields a ground-state or eq-1RDM that equals the true ground-state or eq-1RDM,
\begin{align}\label{eq.def.1rdm.ks}
	\gamma\eq(x,x') = \gamma_{\sss S,eq}(x,x')\ .
\end{align}

Note that a KS system does not exist in RDMFT for Coulombic matter at zero temperature.
The reason behind this is the presence of the electron-electron cusp emerging from
the interelectronic repulsion\cite{K57}. Capturing this cusp
requires a superposition of infinitely many Slater determinants\cite{F03}.
Hence, the gs 1RDM for Coulombic systems has an infinite number of occupied orbitals, 
i.e., natural orbitals with occupation numbers $n_i>0$.  

In the following we show that in FT-RDMFT, however, such a KS system \emph{does indeed exist}.
For a grand canonical ensemble, the eq-1RDM is given by
\begin{align}\label{eq.def.1rdm.ks.2}
\gamma_{\sss S,eq}(x,x') &= \sum_i n_i\, \phi^*_i(x')\, \phi_i(x)\,,
\end{align}
where the natural orbitals of the 1RDM in the KS system 
are by definition identical to the natural orbitals of the true, 
interacting 1RDM defined in Eq.~(\ref{eq.def.1rdm.2}).
The eigenvalues $\{\varepsilon_i\}$ and the chemical potential $\mu$ completely 
determine the occupation numbers $\{n_i\}$ by the relation
\begin{align}\label{eq.n.e}
n_i &= \frac{1}{1+e^{\beta(\varepsilon_i-\mu)}}\,,
\end{align}
which can be inverted to yield
\begin{align}\label{eq.e.n}
\varepsilon_i-\mu&=\frac1\beta\ln\left(\frac{1-n_i}{n_i}\right)\ .
\end{align}

In contrast to the zero-temperature case, it is now possible 
to construct the KS Hamiltonian in the following way: 
The KS Hamiltonian is obtained via its spectral representation in Eq.~(\ref{KS.Ham}), 
where its eigenvalues are determined from Eq.~\Eq{eq.e.n}, 
while its eigenfunctions are given by the natural orbitals of the given 1RDM 
in Eq.~(\ref{eq.def.1rdm.ks.2}). 
The occupation numbers of a KS system in thermal equilibrium at
finite temperature cannot be 0 or 1, as can be seen from Eq.~\Eq{eq.n.e}. 
This is also true for the interacting 1RDM of a grand canonical ensemble, 
as we show in Appendix~\ref{sec.appendix.eq-occ}. 
Hence, it is ensured that the domain of the KS system includes the interacting system.

Furthermore, due to the variational principle the KS potential is generally nonlocal\cite{G75}.
Its uniquess follows from the Hohenberg-Kohn-like theorems shown in Sec.~\ref{sec.FT-RDMFT.hk}. 
It can be expressed explicitly as
\begin{align}\label{eq.veff}
v\s(x,x') &= \sum_{i,j} \left( \delta_{ij} \varepsilon_i - t_{ij} \right) \phi^*_i(x') \phi_j(x)\,,
\end{align}
where $t_{ij}=\bra{\phi_i} \hat T \ket{\phi_j}$ is the kinetic operator in the basis of natural orbitals. 
The requirement of locality can be imposed on the KS potential. This is computationally advantageous, 
but leaves the domain of an exact theory, because this requirement results in approximate natural orbitals 
that cannot be equal to the true natural orbitals\cite{LHRG14}.

Having established the KS scheme, 
we can express the grand potential of the interacting system as 
\begin{align}\label{eq.gp.def}
\Omega[\gamma] &=
F[\gamma] + V[\gamma] -\mu N[\gamma]\,,
\end{align}
where we express the universal functional in terms of common KS quantities as 
\begin{align}\label{FKS.def}
F[\gamma] &= \Omega_k[\gamma] - \frac{S_0[\gamma]}{\beta} 
+ \Omega\h[\gamma] +\Omega\x[\gamma] + \Omega\c[\gamma] .  
\end{align}
Here, 
\begin{align}
\Omega_k[\gamma]&=\int dx'\lim_{x\rightarrow x'}\left(-\frac{\nabla^2}{2}\right)\gamma(x,x')\,,\\
V[\gamma]        &= \int dxdx' v(x,x')\gamma(x',x)\,,\\
N[\gamma]        &= \int dx \gamma(x,x)\,,\\
S\s[\gamma]      &=-\sum_i\left(n_i\ln n_i+(1-n_i)\ln(1-n_i)\right)\,,\\
\Omega\h[\gamma] &= \frac{1}{2} \int dxdx' w(x,x')\gamma(x,x)\gamma(x',x')\,,~\textrm{and}\label{omega_h}\\ 
\Omega\x[\gamma] &= -\frac{1}{2} \int dxdx' w(x,x')\gamma(x,x')\gamma(x',x)\label{omega_x} 
\end{align}
denote the functionals of kinetic energy, external potential, particle number, 
KS entropy, Hartree, and exchange energy, 
which are known explicitly\footnote{In Eq.~(\ref{eq.hf.w}) 
we justify the definitions of the Hartree and exchange energies 
by deriving them explicitly from FT-MBPT.}. 
The remaining term $\Omega\c[\gamma]$ is the correlation contribution,
but its exact form is not known explicitly.

\subsection{Adiabatic connection formula in finite-temperature reduced density matrix functional theory}
\label{sec.FT-RDMFT.ac}

We derive the adiabatic connection formula in FT-RDMFT which
allows us to connect the interacting system to the KS system 
with the same eq-1RDM and forms the basis for systematically constructing 
approximations to the correlation functional $\Omega\c[\gamma]$ via FT-MBPT. 

Closely following the standard zero-temperature DFT approach\cite{LP85,GL93}, 
we begin by introducing a coupling constant $\lambda$ into the electronic Hamiltonian 
\begin{align}\label{ac.hamiltonian}
\hat H^\lambda = \hat T + \lambda \hat W + \hat V^\lambda\,, 
\end{align}
where $0 \le \lambda \le 1$.
The potential $\hat V^\lambda$ is chosen such that for any $\lambda$
there is an associated eq-SDO $\hat D^\lambda[\gamma]$ that leaves 
the eq-1RDM invariant under a change of $\lambda$. 
Along with that we define an auxiliary Hamiltonian
\begin{align}\label{aux.hamiltonian}
\hat H_a^\lambda = \hat T + \lambda \hat W + \hat V\,, 
\end{align}
such that it agrees with Eq.~(\ref{ac.hamiltonian}) at full coupling-strength 
when $\lambda=1$, i.e., $\hat H_a^1 = \hat H^1$. Additionally, we also 
introduce an auxiliary potential $\hat V_a^\lambda= \hat V^\lambda - \hat V$
such that $\hat H_a^\lambda +\hat V_a^\lambda = \hat H^\lambda$.
The grand potential for the auxiliary Hamiltonian becomes
\begin{align}
\Omega_a^\lambda[\gamma] 
&= \min_{\hat D\to\gamma} 
\Tr{\hat D\left(\hat H_a^\lambda -\mu\hat N +\ln\hat D/\beta \right)}\ .
\end{align}
With the aid of the auxiliary potential we obtain
\begin{multline}\label{min.omega_a}
\Omega_a^\lambda[\gamma] = \min_{\hat D\to\gamma}\left( 
\Tr{\hat D\left(\hat H_a^\lambda + \hat V_a^\lambda -\mu\hat N +\ln\hat D/\beta \right)}\right.\\
\left. -\Tr{\hat D \hat V_a^\lambda}\right)\ .
\end{multline}
Since $\hat V_a^\lambda$ is a one-particle operator, 
we can take the last term out of the minimization and replace $\hat D$ in this term
by any $\hat D\to\gamma$. 
Then, the minimization in Eq.~(\ref{min.omega_a}) yields 
the eq-SDO $\hat D^\lambda = \exp{\hat H^\lambda -\mu\hat N}/\Tr{\exp{\hat H^\lambda}}$ that 
is associated with $\hat H^\lambda$ yielding the eq-1RDM of the true interacting system 
that is invariant under a change of $\lambda$.
Hence, the grand potential becomes
\begin{multline}\label{omega_a}
\Omega_a^\lambda[\gamma] =  
\Tr{\hat D^\lambda\left(\hat H_a^\lambda + \hat V_a^\lambda -\mu\hat N +\ln\hat D^\lambda\beta \right)}\\
-\Tr{\hat D \hat V_a^\lambda}\ .
\end{multline}

By definition $\Omega_a^1[\gamma] = \Omega[\gamma]$, i.e.,  
the auxiliary grand potential at full coupling strength is 
identical to the true interacting grand potential, 
therefore 
\begin{align}\label{gp.1}
\Omega[\gamma] 
&= \Omega_a^0 + \int_0^1 d\lambda\frac{d\Omega_a^\lambda[\gamma]}{d\lambda}\ . 
\end{align}
Taking the derivative with respect to the coupling constant is simplified by the fact
that we consider a system in thermal equilibrium.
Hence, only $\lambda\hat W$ and $\hat V_a^\lambda$ 
contribute to the coupling-constant derivative in Eq.~\Eq{gp.1}, 
yielding
\begin{align}\label{gp.2}
\Omega[\gamma] 
&= \Omega_a^0 + \int_0^1 d\lambda \Tr{\hat D^\lambda \hat W}\ . 
\end{align}
Consider the grand potentials
\begin{align}
\Omega[\gamma] &= \Tr{\hat D\left(\hat T + \hat V +\hat W -\mu\hat N + \ln\hat D/\beta \right)}\,, \\ 
\Omega_a^0[\gamma]&= \Tr{\hat D^0\left(\hat T + \hat V -\mu\hat N + \ln\hat D^0/\beta \right)}
\end{align}
and take into account that $\hat D$ and $\hat D^0$ yield the same eq-1RDM, 
hence, the same expectation values of one-particle operators, such as
$\Tr{\hat D \hat T} = \Tr{\hat D^0 \hat T}$,
$\Tr{\hat D \hat V} = \Tr{\hat D^0 \hat V}$, and
$\Tr{\hat D \hat N} = \Tr{\hat D^0 \hat N}$.

Then, we can further reduce Eq.~(\ref{gp.2}) and obtain the
adiabatic connection formula for the entire interaction as
\begin{align}\label{eq.gp_final}
\Omega\w[\gamma] &= \int_0^1 d\lambda W^\lambda[\gamma]\,,
\end{align}
where we define $\Omega\w = W-S\c/\beta$, $W^\lambda=\Tr{\hat D^\lambda \hat W}$,
$W=\Tr{\hat D \hat W}$ and 
$S\c=\Tr{\hat D^0 \ln\hat D^0} - \Tr{\hat D -\ln\hat D}$ 
as the entropic correlation contribution.

Finally, by subtracting the Hartree and exchange contributions
defined as $W\hx = \Tr{\hat D^0 \hat W}$ we obtain the 
adiabatic connection formula for the correlation contribution
\begin{align}\label{ac.omega_c}
\Omega\c[\gamma] 
&=  \int_0^1\frac{d\lambda}{\lambda} W\c^\lambda[\gamma]\,,
\end{align}
where we define $\Omega\c = W\c - S\c/\beta$, $W\c= W-W\hx$,
and $W\c^\lambda = \Tr{\lambda (\hat D^\lambda - \hat D^0)\hat W}$. 

In analogy to DFT, Eq.~(\ref{ac.omega_c}) allows us to express the correlation contribution  
to the KS system in FT-RDMFT as a contribution coming solely from the interaction potential. 
It is interesting to note another similarity between DFT and FT-RDMFT. 
In DFT, the adiabatic connection formula includes the kinetic correlation contribution, 
i.e., the difference between the kinetic energy of the interacting system and the KS 
system via the coupling constant integration. In FT-RDMFT, 
where there is no kinetic correlation contribution, the coupling-constant integral 
instead incorporates the entropic correlation contribution $S\c$. 

The adiabatic connection formulas derived in Eqs.~(\ref{eq.gp_final}) and 
(\ref{ac.omega_c}) are central results, because they are our key 
for developing systematic approximations to $\Omega\w[\gamma]$ 
and its correlation contribution $\Omega\c[\gamma]$ based on FT-MBPT.

\subsection{Constructing correlation functionals}
\label{sec.FT-RDMFT.functionals}

With the aid of the adiabatic connection we can use methods from 
FT-MBPT\cite{FW03} to systematically construct approximations 
to the functionals $\Omega\w[\gamma]$ and $\Omega\c[\gamma]$, 
where the KS system, defined by the Hamiltonian $\hat H^0$, 
serves as our reference system in a perturbative expansion.
Our starting point from the perspective of FT-MBPT 
is to relate the temperature Green's function $\mathcal{G}^\lambda(x,\tau,x',\tau')$ 
to the adiabatic connection in Eq.(\ref{eq.gp_final}).
This relation is expressed as
\begin{multline}\label{ft-mbptgp}
\lambda W^\lambda 
=\frac12\int dxdx'\lim_{\tau'\rightarrow\tau^+}\Bigg(-\delta(x-x')\frac{\partial}{\partial\tau}\\
-\Big(-\delta(x-x')\frac{\nabla^2}{2}+v\s(x',x)\Big)\Bigg)\mathcal{G}^\lambda(x,\tau,x',\tau')\\
+\frac12\int dxdx' (v\s(x',x)-v^\lambda(x',x))\gamma(x,x')\,,
\end{multline}
where $\tau=i t$ denotes imaginary time 
and $\tau^+ = \lim_{\eta\to 0^+} (\tau + \eta)$.
The use of FT-MBPT in Eq.~(\ref{ft-mbptgp}) is facilitated 
by the existence of the adiabatic connection which 
connects the true interacting system with the KS system,
and hence,  allows to express the resulting Feynman diagrams 
in terms of occupation numbers and natural orbitals of the 1RDM.

Well-known methods of FT-MBPT can now be applied.
The unperturbed Hamiltonian is $\hat H^0$, whereas 
the peturbation consists of 
a two-particle interaction $\lambda\hat W$ and 
a nonlocal one-particle potential $u^\lambda(x,x')=v^\lambda(x,x') -v\s(x,x')$.
The proof of Wicks theorem is still applicable for this kind of perturbation and the same
Feynman rules apply. We show our notation conventions in Table \ref{tab.feyn}.
\setlength{\unitlength}{1mm}
\setlength{\ltiny}{5mm}
\setlength{\lsmall}{10mm}
\setlength{\lmedium}{15mm}
\setlength{\lbig}{20mm}
\setlength{\lbigg}{30mm}
\setlength{\lhuge}{40mm}
\renewcommand{\xtiny}{5}
\renewcommand{\xsmall}{10}
\renewcommand{\xbig}{20}
\renewcommand{\xbigg}{30}
\renewcommand{\xhuge}{40}
\renewcommand{\arrowshort}{7}
\renewcommand{\arrowlong}{10}
\renewcommand{\arrowang}{25}
\begin{fmffile}{gp.one}
\fmfset{dot_len}{1.5mm}
\fmfset{dash_len}{2.5mm,1mm}
\fmfset{arrow_len}{\arrowshort}
\fmfset{arrow_ang}{\arrowang}
%
% \subsubsection{Feynman rules}
%
%
%
% \begin{minipage}{.5\textwidth}
% \begin{tabular}[c]{|l|l|}
\begin{table}[t]
  \begin{center}
  \begin{tabular}[c]{|ll|}
    \hline&\\
      \begin{minipage}[c][\ltiny][c]{\lbig}
	\begin{center} 
	   \includegraphics{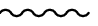}
% 	  \begin{fmfgraph}(\xmedium,\xbig)
% 	      \fmfleft{i}\fmfright{o}
% 	      \fmf{wiggly}{i,o}
% 	  \end{fmfgraph}
	  \end{center}
	\end{minipage} &
	  \begin{minipage}[l]{40mm}
	      $\lambda w(x,x')$ 
	  \end{minipage}\\
    &\\
  %    \hline&\\
		  \begin{minipage}[c][\ltiny][c]{\lbig}
	  \begin{center} 
% 			  \begin{fmfgraph}(\xmedium,\xbig)
% 	      \fmfleft{i}\fmfright{o}
% 	      \fmf{zigzag}{i,o}
% 	  \end{fmfgraph}
	    \includegraphics{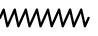}
	  \end{center}
	\end{minipage} &
	  \begin{minipage}[l]{40mm}
	      $u^\lambda(x,x')$ 
	  \end{minipage}\\
    &\\
  %    \hline&\\
		  \begin{minipage}[c][\ltiny][c]{\lbig}
	  \begin{center} 
% 			  \begin{fmfgraph}(\xmedium,\xbig)
% 	      \fmfleft{i}\fmfright{o}
% 	      \fmf{plain_arrow}{i,o}
% 	  \end{fmfgraph}
	    \includegraphics{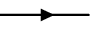}
	  \end{center}
	\end{minipage} &\begin{minipage}[l]{40mm}
	      $\mathcal{G}^0(x,\tau,x',\tau')$ 
	  \end{minipage}\\
    &\\
  %    \hline&\\
		  \begin{minipage}[c][\ltiny][c]{\lbig}
	  \begin{center} 
% 			  \begin{fmfgraph}(\xmedium,\xbig)
% 	      \fmfleft{i}\fmfright{o}
% 	      \fmf{dbl_plain_arrow}{i,o}
% 	  \end{fmfgraph}
	    \includegraphics{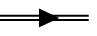}
	  \end{center}
	\end{minipage} &\begin{minipage}[l]{40mm}
	      $\mathcal{G}^\lambda(x,\tau,x',\tau')$ 
	  \end{minipage}\\
    &\\
    \hline
  \end{tabular}
  % \end{minipage}
  \end{center}
  \caption{Notation conventions for Feynman diagrams in FT-RDMFT, 
  where $w(x,x')$ denotes the interelectronic interaction, 
  $u^\lambda(x,x')=v^\lambda(x,x') -v\s(x,x')$ the nonlocal one-particle potential,
  $\mathcal{G}^0(x,\tau,x',\tau')$ the Green's function of the unperturbed system, and
  $\mathcal{G}^\lambda(x,\tau,x',\tau')$ the temperature Green's functions.
  }
  \label{tab.feyn}
\end{table}
 
In particular, if the Hamiltonian is temperature-independent 
and the system is uniform, Eq.~\Eq{gp.2} can be written
entirely in terms of Feynman diagrams as
\begin{align}
  \lambda W^\lambda_{unif}&=\frac12\left(\hspace{-4mm}
  \begin{minipage}{60pt}
%     \fbox{
%     \begin{fmfgraph*}(30,10)
%       \fmfbottom{i1}
%       \fmftop{o1}
%       \fmf{\phantom}{i1,v1}
%       \fmf{\phantom,tension=.5}{v1,v2,v3}
%       \fmf{\phantom}{v3,o1}
%       \fmffreeze
%       \fmf{dots,left}{v1,v3,v1}
%       \fmf{dbl_plain_arrow,right=2.5}{v3,v1}
%       \fmfv{label=$\Sigma^*$,label.dist=-4,decor.size=20,decor.shape=circle,decor.filled=empty}{v2}
%     \end{fmfgraph*}
%     }
    \includegraphics{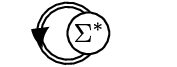}
  \end{minipage}
  -
  \begin{minipage}[c]{50pt}
      \mbox{
%       \begin{fmfgraph}(\xmedium,\xmedium)
%         \fmfleft{i1}
%         \fmfright{o1}
%         \fmf{zigzag}{i1,o1}
% 	\fmfdot{i1,o1}
% %           \fmf{phantom}{i2,o2} 
% %           \fmf{wiggly,right=.7}{i1,o1}
% %           \fmfdot{i1,o1}
%         \fmf{dbl_plain_arrow,left=.7}{i1,o1}
%       \end{fmfgraph}
	\includegraphics{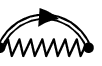}
      }
    \end{minipage}\label{eq.lambda_w}
    \right)\,,
\end{align}
where $\Sigma^*$ denotes the irreducible self-energy.
Whereas in general, the irreducible self energy $\Sigma^*$ 
for the first-order contribution becomes
\begin{align}
  \Sigma^*&=
    \hspace{1mm}
    \begin{minipage}[c]{\lsmall}
      \mbox{
%       \begin{fmfgraph}(\xtiny,\xmedium)
%         \fmfbottom{i1,i2}\fmftop{o1,o2}
%           \fmf{fermion}{i1,o1}
%           \fmf{phantom}{i2,o2} 
%           \fmf{wiggly,right=.7}{i1,o1}
%           \fmfdot{i1,o1}
%       \end{fmfgraph}
      \includegraphics{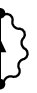}
      }
    \end{minipage}
    \hspace{-2mm}
  +
    \hspace{1mm}
    \begin{minipage}[c]{\lbig}
      \mbox{
%       \begin{fmfgraph}(\xmedium,\xbig)
%         \fmfleft{i1}
%         \fmfright{o1}
%         \fmf{wiggly,tension=2}{i1,v1}
%         \fmf{plain_arrow,left}{v1,o1,v1}
%         \fmfdot{i1}
%       \end{fmfgraph}
	\includegraphics{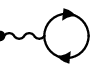}
      }
    \end{minipage}
    \hspace{-1mm}
  +
    \hspace{2mm}
    \begin{minipage}[c]{\lsmall}
      \mbox{
%       \begin{fmfgraph}(\xtiny,\xmedium)
%         \fmfbottom{i1,i2}\fmftop{o1,o2}
%           \fmf{zigzag}{i1,o1}
%           \fmf{phantom}{i2,o2} 
%           \fmfdot{i1,o1}
%       \end{fmfgraph}
      \includegraphics{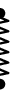}
      }
    \end{minipage}\label{eq.1st}.
\end{align}
\end{fmffile}
Combining Eqs.~\Eq{eq.gp_final}, \Eq{gp.2} and \Eq{eq.1st}, we arrive at the first-order
contribution to the interaction-induced grand potential functional in FT-RDMFT which is
\begin{align}
\Omega^{(1)}\w[\gamma]&=\Omega\h[\gamma]+\Omega\x[\gamma]\label{eq.hf.w}\,,\textrm{~where~}\\
\Omega\h[\gamma]&=\frac12\int dxdx' w(x,x')\gamma(x,x)\gamma(x',x')\label{eq.hf.h}\textrm{~and~}\\
\Omega\x[\gamma]&=-\frac12\int dxdx' w(x,x')\gamma(x,x')\gamma(x',x)\label{eq.hf}.
\end{align}
This justifies the definitions of the Hartree and exchange energies which we postulated
in Eqs.~(\ref{omega_h}) and (\ref{omega_x}).
Note that the functional form of the first-order contributions are equivalent to the
Hartree and exchange functionals in zero-temperature RDMFT\footnote{We show 
in Appendix~\ref{sec.appendix.FT-HF} that it is appropriate to 
call $\Omega^{(1)}\w[\gamma]$ Hartree-Fock functional with 
$\Omega\h[\gamma]$ and $\Omega\x[\gamma]$ representing Hartree and exchange contributions,
because the minimization of the grand potential using the functionals
\Eq{eq.hf.w}-\Eq{eq.hf} is equivalent to the solution 
of the finite-temperature Hartree-Fock equations.}.

Approximations for the correlation functional $\Omega\c = \Omega\w-\Omega^{(1)}\w$
can now be derived by expanding the Green's function to higher orders 
and then solving Eqs.~\Eq{eq.gp_final} and \Eq{ft-mbptgp}.

\section{Summary and conclusions}
In this work, we have derived and presented the foundations of FT-RDMFT.
We have proven Hohenberg-Kohn-like theorems and 
shown that the equilibrium properties of a grand canonical ensemble
with nonlocal external potential are determined uniquely by the eq-1RDM. 
This allows us to establish a functional theory for the grand potential in terms of
the 1RDM and, in analogy to DFT, to define a universal functional. 
A minimization of that grand potential functional then yields the eq-1RDM. 

Furthermore, we have shown that there exists a KS system in FT-RDMFT, 
in contrast to the zero-temperature case, 
and derived the adiabatic connection formula. 
Based on this, we have established an iterative procedure for
constructing approximations to the correlation functional in FT-RDMFT by utilizing 
methods from FT-MBPT. We have further demonstrated that the minimization of 
the first-order functional in this perturbative scheme is equivalent to the solution of
the finite-temperature Hartree-Fock equations.

The present work sparks the hope that FT-RDMFT might become 
the method of choice for quantum problems at finite temperature 
where the standard DFT approach fails 
and the thermal DFT approach has not been developed to satisfaction\cite{PPGB14}. 

The main task for the future is the development of correlation functionals for the grand potential and
free energy in FT-RDMFT and the application to real systems. 
Some further developments, such as exchange-only functional for collinear and non-collinear spin,
as well as correlation functionals, momentum distributions, and phase diagrams 
in the framework of FT-RDMFT will be presented in Ref.~\cite{BECG15}. 

\appendix
\section{Equilibrium occupation numbers in general systems}
\label{sec.appendix.eq-occ}

As we have pointed out in Section \ref{sec.FT-RDMFT.ks}, the eq-1RDM of a noninteracting system
has occupation numbers strictly between 0 and 1. We now show that this is also
true for the occupation numbers of eq-1RDMs of arbitrary systems, including interacting
ones.

We start from the spectral representation of the eq-1RDM given by
\begin{align}
  \gamma(x,x')&=\sum_in_i\phi_i^*(x')\phi_i(x)\ .
\end{align}
The occupation number operator $\hat n_i$ is now defined as
\begin{align}
  \hat n_i&=\hat c_i^+\hat c_i\,,
\end{align}
where $c_i^+$ creates and $c_i$ annihilates the natural orbital $\phi_i$.
An arbitrary occupation number of the eq-1RDM in grand canonical equilibrium can then be written as
\begin{align}
  n_i&=\tr{\hat D\hat n_i}=\sum_ew_e\bra{\Psi_e}\hat n_i\ket{\Psi_e}\label{eq.app.ni.inf}.
\end{align}
The $\{\Psi_e\}$ are eigenfunctions of the Hamiltonian and form a basis of
the underlying Hilbert space. Another basis is formed by the Slater determinants
$\{\Phi_\alpha\}$ which are constructed by the natural orbitals $\{\phi_i\}$ of the
eq-1RDM. The transformation between these bases is governed by the expansion
coefficients $c_{e\alpha}$ via
\begin{align}
  \Psi_e&=\sum_\alpha c_{e\alpha}\Phi_\alpha.
\end{align}
Due to completeness and normalization of the $\{\Psi_e\}$ and $\{\Phi_\alpha\}$, the coefficients fulfill
\begin{align}
  \sum_e|c_{e\alpha}|^2&=\sum_\alpha|c_{e\alpha}|^2=1\label{eq.app.norm}.
\end{align}
Expanding the $\{\Psi_e\}$ in Eq.~\Eq{eq.app.ni.inf} in terms of the $\{\Phi_\alpha\}$ then leads to
\begin{align}
  n_i&=\sum_ew_e\sum_{\alpha\beta}c_{e\alpha}^*c_{e\beta}\bra{\Phi_\alpha}\hat n_i\ket{\Phi_\beta}\ .
\end{align}
Since the Slater determinants $\{\Phi_\alpha\}$ are by definition eigenfunctions of
the occupation number operator $\hat n_i$, this reduces to
\begin{align}\label{eq.app.ni_final}
  n_i&=\sum_\alpha\underbrace{\left(\sum_ew_e|c_{e\alpha}|^2\right)}_{f_\alpha}\underbrace{\bra{\Phi_\alpha}\hat n_i\ket{\Phi_\alpha}}_{g_{i\alpha}}.
\end{align}
Using Eq.~\Eq{eq.app.norm} and the properties of the thermal weights, $w_e>0$ and $\sum_ew_e=1$, we see that
\begin{align}
  f_\alpha&>0\label{eq.app.f_prop_1}\\
  \sum_\alpha f_\alpha&=1\label{eq.app.f_prop_2}.
\end{align}
The factors $g_{i\alpha}$ are equal to 1, if the natural orbital $\phi_i$ appears in the
Slater determinant $\Phi_\alpha$. Otherwise $g_{i\alpha}$ vanishes. 
The summation over $\alpha$ corresponds
to a summation over a basis of the Hilbert space, which is the Fock space in the case of a grand
canonical ensemble. Therefore, for a fixed $i$, there will be at least one
$\alpha$, such that $g_{i\alpha}=1$ and at least one $\alpha$ for which  $g_{i\alpha}=0$.
Combining this fact with Eqs.~\Eq{eq.app.f_prop_1} and \Eq{eq.app.f_prop_2},
we can rewrite Eq.~\Eq{eq.app.ni_final} to yield the desired inequality
\begin{align}
  0&<n_i<1 .
\end{align}

\section{Zero-temperature mapping between potentials and wavefunctions}
\label{sec.appendix.counter}
Assume an arbitrary Hamiltonian 
\begin{align}
  \hat H&=\hat T+\hat V+\hat W\\
\end{align}
with gs-1RDM 
\begin{align}
  \gamma_{gs}(x,x')&=\sum_in_i\phi^*_i(x')\phi_i(x)\ .
\end{align}
Due to Gilbert's theorem\cite{G75}, 
the wavefunction can be written as a functional of the 1RDM 
allowing us to define an energy functional
\begin{align}
  E[\gamma]&=\bra{\Psi[\gamma]}\hat H\ket{\Psi[\gamma]}\ .
\end{align}
From the variational principle we infer that this functional is minimal for the gs-1RDM.
Therefore the following relations hold
\begin{align}
  \left.\frac{\partial E[\gamma]}{\partial n_i}\right|_{\gamma_{gs}}&=\left\{
    \begin{array}{ll}
      a_i>\mu&,n_i=0\\
      \mu&,0<n_i<1\\
      b_i<\mu&,n_i=1
    \end{array}
  \right.\,,
\end{align}
where we identify $\mu$ as the chemical potential of the system. 

In the following we show that there is a one-to-one mapping between potential and gs-1RDM, 
if and only if there are no pinned occupation numbers, i.e., occupation numbers equal to 0 or 1. 
This is done in two steps: 
(i) For unpinned occupation numbers we show that the
external potential is uniquely determined up to a constant; 
(ii) considering gs-1RDMs with pinned occupation numbers we show that 
one can explicitly construct infinitely many potentials which leave the gs-1RDM invariant. 

\paragraph*{\textup{\textbf{(i) Unpinned states~ ~}}}
The absence of pinned states allows us to use the following
Euler-Lagrange equation
\begin{align}\label{eq.app.b.euler}
\left.\frac{\delta E[\gamma]}{\delta\gamma(x,x')}\right|_{\gamma_{gs}}
&=\mu\left.\frac{\delta N[\gamma]}{\delta\gamma(x,x')}\right|_{\gamma_{gs}}=\mu\delta(x,x')\,,
\end{align}
where $\delta(x,x')$ denotes the Dirac delta function.
Adding an arbitrary potential contribution $U[\gamma]$ to the energy functional
$E_u[\gamma]=E[\gamma]+U[\gamma]$ then yields
\begin{align}
  \left.\frac{\delta E_u[\gamma]}{\delta\gamma(x,x')}\right|_{\gamma_{gs}}&=\mu\delta(x,x')+u(x,x'). 
\end{align}
As every gs-1RDM has to fulfill Eq.~\Eq{eq.app.b.euler}
we deduce that the only choice of $u(x,x')$ 
that leaves the gs-1RDM invariant is
\begin{align}
  u(x,x')&=c\delta(x,x')
\end{align}
with $c$ being an arbitrary constant.

\paragraph*{\textup{\textbf{(ii)Pinned states~ ~}}}
For pinned occupation numbers the minimum of
$E[\gamma]$ is at the boundary of the domain, 
and hence, we cannot use Eq.~\Eq{eq.app.b.euler}.
It is possible to adjust the Euler-Lagrange equation by incorporating
Kuhn-Tucker multipliers\cite{KT51}, but there is a simpler way, as described
in the following.

We exploit the fact that the derivatives in Eq.~\Eq{eq.app.b.euler} 
can be different from $\mu$ for pinned states
and construct a one-particle potential which leaves the gs-1RDM invariant. 
This potential shall be governed by the generally nonlocal kernel $u(x,x')$. By
choosing it diagonal in the natural orbital basis of the gs-1RDM we ensure that
the orbitals do not change upon addition of the potential. For simplicity, we choose
only one component to be non-vanishing, namely
\begin{align}\label{eq.app.b.u}
  u(x,x')&=u\phi_\alpha^*(x')\phi_\alpha(x)\ .
\end{align}
We then define an energy functional  
\begin{align}
  E_\alpha[\gamma]&=E[\gamma]+\int dxdx'u(x,x')\gamma(x',x)
\end{align}
for which the derivative with respect to the occupation numbers becomes
\begin{align}\label{eq.app.b.arg}
  \left.\frac{\partial E_\alpha[\gamma]}{\partial n_i}\right|_{\gamma_{gs}}&=\left\{
    \begin{array}{ll}
      a_i+u\delta_{i\alpha}&,n_i=0\\
      \mu+u\delta_{i\alpha}&,0<n_i<1\\
      b_i+u\delta_{i\alpha}&,n_i=1
    \end{array}
  \right.\ .
\end{align}

These considerations can now be employed to show the ambiguity of the external potential
in RDMFT for groundstates with pinned occupation numbers. For simplicity we
assume that there is exactly one pinned occupation number with $n_\beta=0$. We then
construct a potential of the form given in Eq.~\Eq{eq.app.b.u} with $\alpha=\beta$. 
We deduce from Eq.~\Eq{eq.app.b.arg} that every choice of $u>\mu-a_\beta$ 
leads to a situation where the $\beta$-orbital exhibits a derivative bigger than $\mu$, 
leaving the gs-1RDM invariant. For one pinned ocuppation number $n_\beta=1$ we can choose $u<\mu-b_\beta$
which lets the derivative of the $\beta$-orbital always surpass $\mu$ again leading to
the same gs-1RDM. When considering gs-1RDM with several pinned states these arguments are
readily generalized. This proofs the ambiguity of the one-particle potential for gs-1RDM
with pinned occupation numbers.

\section{Finite-temperature Hartree-Fock theory}
\label{sec.appendix.FT-HF}

We consider the functional in Eq.~\Eq{eq.gp.def}, but without the correlation contribution
\begin{multline}\label{eq.gp_hf}
  \Omega_{HF}[\gamma]=\Omega_k[\gamma]+V[\gamma]-\mu N[\gamma]- S\s[\gamma]/\beta\\
  +\Omega\h[\gamma]+\Omega_x[\gamma].
\end{multline}
The derivative of the KS entropy $S\s[\gamma]$ with respect to the occupation numbers 
diverges for $n_i\rightarrow\{0,1\}$ whereas all other contributions are finite.
Therefore, there are no pinned states at the minimum of $\Omega_{HF}[\gamma]$. 
Furthermore, Eq.~\Eq{eq.gp_hf} is an explicit functional of the 1RDM. 
Therefore we conclude that the functional derivative with respect to the 1RDM exists 
and that $\Omega_{HF}[\gamma]$ fulfills the Euler-Lagrange equation 
\begin{align}
  \frac{\delta \Omega_{HF}[\gamma]}{\delta \gamma(x',x)}&=0
\end{align}
at the minimum. 
Applying this condition on $\Omega_{HF}[\gamma]$ and
projecting the result on the $i$-th natural orbital of the 1RDM
leads to the FT-HF equations
\begin{align}
  0=&\int dx'\phi_{i}(x')\frac{\delta \Omega_{HF}[\gamma]}{\delta \gamma(x',x)}\\
  =&\left(-\frac{\nabla^2}{2}\right)\phi_{i}(x)+\int dx'v(x,x')\phi_{i}(x')\nn\\
  &\hspace*{5mm}-\int dx'w(x,x')\gamma(x,x')\phi_{i}(x')\nn\\
  &\hspace*{12mm}+\int dx'w(x,x')\gamma(x',x')\phi_{i}(x)-\veps_{i}\phi_{i}\label{eq.eq.ftrdmft.fthf},
\end{align}
where we used Eq.~\Eq{eq.e.n} in the last term. Hence, it is appropriate 
to call Eq.~(\ref{eq.gp_hf}) Hartree-Fock functional.

\section{Canonical ensembles}
\label{sec.appendix.canonical}

Minimizing the grand potential implies the coupling to a particle bath. 
There are, however, important physical problems
where the particle number is intrinsically conserved and therefore the equilibrium is
defined by the state which minimizes the free energy instead. 
An important example is a molecule in solution where the solvent is described by a
bath of harmonic oscillators at finite temperature. 
In the following we present a concise formulation of FT-RDMFT for canonical ensembles
and point out the most important differences to the grand canonical formulation.

The appropriate Hilbert space for canonical ensembles of particle
number $N$ is given by the $N$-particle subspace
\begin{align}
\mathcal{H}^N&=\hat Sh^{\otimes N},
\end{align}
of the Fock space 
\begin{align}
\mathcal{H}=\bigoplus_{n=0}^{\infty}\hat Sh^{\otimes n}.
\end{align}
The associated SDOs are weighted sums of projection operators on $\mathcal{H}^N$:
\begin{multline}
  \hat D^{c}=\sum_{\alpha}w_{\alpha N}\ket{\Psi_{\alpha N}}\bra{\Psi_{\alpha N}},\\w_{\alpha N}\geq0,\sum_{\alpha,}w_{\alpha N}=1\label{eq.sdo.c}.
\end{multline}
The variational principle now involves the free energy
$\mathcal{F}=\Tr{\hat D\left(\hat H + 1/\beta\ln\hat D \right)}$ 
rather than the grand potential and the eq-SDO is given by
\begin{align}
  \hat D^c\eq&=\frac{e^{-\beta\hat H}}{\tr{e^{-\beta\hat H}}},
\end{align}
where $\hat H$ is now the $N$-particle Hamiltonian of the system.
The one-to-one mapping between the eq-SDO and the eq-1RDM stays
valid also in the case of canonical ensembles with the only difference that the external
potential is now determined only up to an additional constant. 
Following the construction by Lieb\cite{L83} we define a canonical universal functional 
$\mathfrak{F}^c[\gamma]$ on the whole domain of ensemble-N-representable 1RDMs as
\begin{align}
\mathfrak{F}^c[\gamma] 
&=\inf_{\hat D\in\mathcal{H}^N\rightarrow\gamma}\tr{\hat D(\hat T+\hat W+1/\beta\ln\hat D)}.
\end{align}
The equilibrium of the system is then found by a minimization of the free energy
functional $\mathcal{F}[\gamma]=\mathfrak{F}^c[\gamma]+V[\gamma]$.

The main difference to the grand canonical framework of FT-RDMFT lies in 
the canonical KS system.
In the canonical ensemble a simple analytical relation between the eigenvalues 
of the KS Hamiltonian and the occupation numbers as in Eq.~\Eq{eq.n.e} for the canonical ensemble
does not exist. Therefore we do not know if every 1RDM with $0<n_i<1$ is a canonical eq-1RDM. 
Nevertheless, we can reconstruct the KS Hamiltonian by iterative methods\cite{WP93}
once we know that a given 1RDM corresponds to a canonical equilibrium.

Furthermore, the finite temperature version of Wick's theorem\cite{M55} breaks down 
for canonical ensembles, because it relies on the interplay of
states of different particle numbers.
Hence, our perturbative approach for constructing approximations to the correlation 
functional cannot be applied to the canonical ensemble in general. 
However, there is a loophole. When we consider the system in the thermodynamic limit, 
the thermodynamic variables of grand canonical and canonical ensembles coincide. 
In this case, we can still use functionals derived by our perturbative methodology 
of Sec.~\ref{sec.FT-RDMFT.functionals} 
for the grand potential $\Omega[\gamma]$ and calculate the free energy via
\begin{align}
\mathcal{F}[\gamma]&=\Omega[\gamma]+\mu N[\gamma].
\end{align}

%\begin{acknowledgments}
%\end{acknowledgments}

\bibliography{master_AC}

\end{document}